# Gyrator-Capacitor Modeling of A Continuously Variable Series Reactor in Different Operating Modes

Mohammadali Hayerikhiyavi
Department of Electrical and Computer Engineering
University of Central Florida
Orlando, USA
Mohammad.ali.hayeri@knights.ucf.edu

Aleksandar Dimitrovski
Department of Electrical and Computer Engineering
University of Central Florida
Orlando, USA
ADimitrovski@ ucf.edu

***Abstract*— Continuously Variable Series Reactor (CVSR) can regulate its reactance on the ac side using the nonlinear ferromagnetic core, shared by an ac and a dc winding for applications like power flow control, oscillation damping, load balancing and others. In order to use a CVSR in the power grid, it is essential to know its operating characteristics. The Gyrator-Capacitor approach has been applied to model the magnetic circuit coupled with the electrical circuit. In the paper, we investigate the CVSR behavior in terms of induced voltages across the dc winding, flux densities (B) through the core legs, and power exchange between the ac controlled and the dc control circuit.**

## I. Introduction

The existing power grids are already under considerable stress and strain, but as the demand for electricity continuously grows and the paradigms of operation are changing, they become even more so. One of the most important challenges power system operators face is the prevention of blackouts due to generation and demand volatility, transmission congestions, oscillations, etc. Most of these problems could be alleviated with a comprehensive ac power flow control [1,2]. Typical devices for power flow control include phase shifting transformers, switched shunt-capacitors/inductors, and various types of flexible ac transmission systems (FACTS) controllers. These devices either can provide only a coarse control or they have high investment and operation costs. Saturable reactor technology does not have these drawbacks and it provides alternative option for power flow control with high reliability at low costs [3,4]. Continuously Variable Series Reactor (CVSR) is a series saturable reactor that has variable reactance within the design limits. Continuous and smooth control of a large amount of power flow on the ac side can be easily achieved by changing the relatively small dc bias current. The dc source for the CVSR is low voltage/current power electronics-based converter. On the other hand, FACTS controllers use high voltage and current components because they are also part of the primary power circuit [5-7]. Besides power flow control, there are other applications that the CVSR is suitable for, such as oscillation damping and fault currents limiting [8].

It can add additional impedance into the ac power circuit either to damp oscillations or decrease the fault current when needed. In order to evaluate its functionality, it is necessary to study the CVSR under all system conditions and understand its impact on the power system, as well as, the ways the power system affects its operation.

The basic concept of the CVSR is briefly reviewed in Section II. The gyrator-capacitor (G-C) approach used for the modelling of the magnetic circuit is explained in Section III. Section IV presents simulation results and analysis of the CVSR induced voltage, the power transferred into its dc windings and its equivalent inductance for different DC and AC voltage sources. Conclusions are summarized in Section V.

## II. CVSR

In the saturable-core reactor shown in Figure 1, an ac winding is wound on the middle leg and connected in series with the ac circuit that transfers power from the source to the load. Typically, there is an air gap in the middle leg, necessary to achieve the desired reactance in normal conditions and not to saturate the core with the ac flux. Two dc windings connected in series and controlled by the dc source are wound on the two outer legs. The ac reactance of the reactor is controlled by the dc current [9,10]. This reactance reaches maximum value when the core is not saturated and minimum value when it is fully saturated (at large enough dc current). Thus, the overall reactance in the controlled ac circuit is changed by the dc.

During each half cycle of a period, the ac and dc fluxes will add in one half of the core and subtract in the other half. It means that the reluctance of one leg will decrease and increase in the second [11], when the core operates around the knee of the B-H curve. Therefore, $\frac{d\Phi_{right}}{dt}$ and $\frac{d\Phi_{left}}{dt}$ will be different, and the induced voltages on the right ($V_{right}$) and left dc winding ($V_{left}$) will also be different and impose $V_{bias} = V_{right} - V_{left}$ in the dc circuit. It will be shown in the sequel that the frequency of this voltage is twice the frequency of the ac circuit.

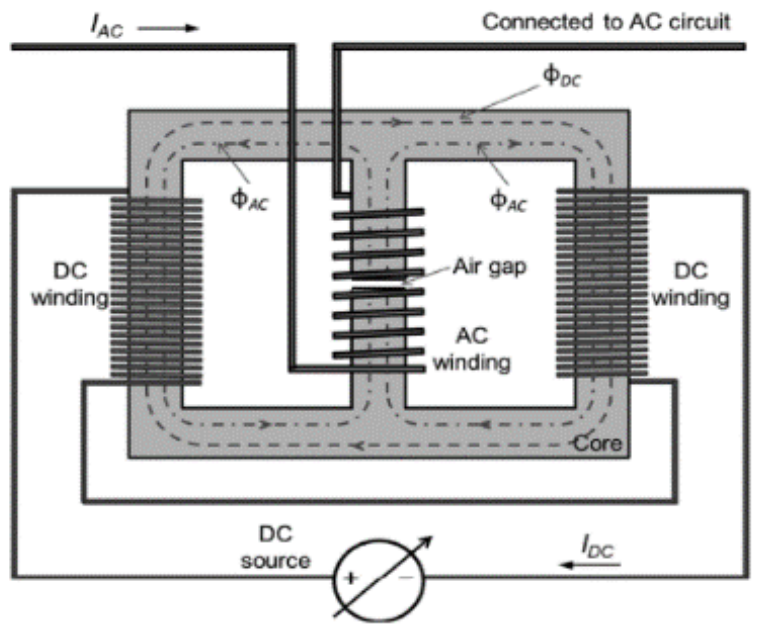

Fig. 1. Schematic of a CVSR [1].

## III. Gyrator-Capacitor model

A typical approach to modelling magnetic circuits is to use an electric circuit analogy [12, 13]. Equivalent circuits are formed using resistors to represent reluctances and voltage sources to represent magnetomotive forces (MMFs). However, magnetic circuits store energy and, strictly speaking, are not properly modelled by resistors which only dissipate energy. The G-C model shown in Figure 2 [14] keeps the power equivalence.

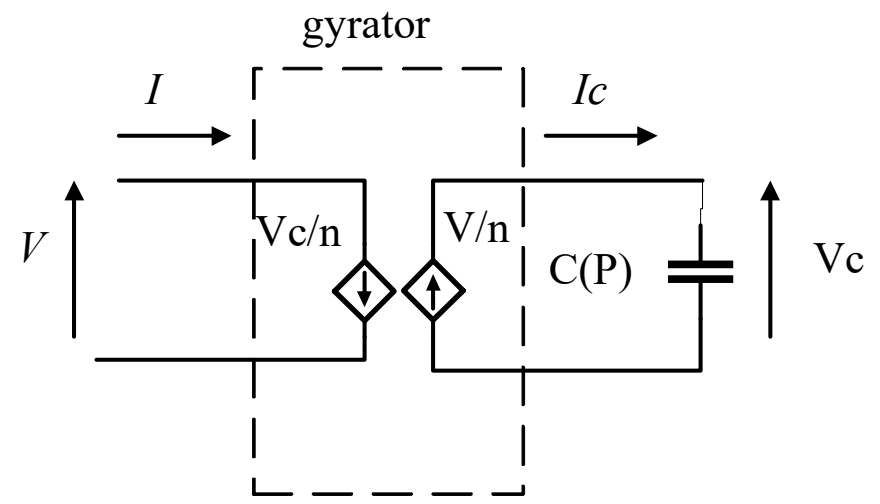

Fig. 2. Gyrator-Capacitor model of a simple magnetic circuit.

In the G-C model, the analogy between MMF and voltage is kept, but electrical current is analogous to the rate-of-change of the magnetic flux $\frac{d\Phi}{dt}$, as described by (1) and (2):

$$V_c \equiv mmf \quad (1)$$

$$I_c \equiv \frac{d\Phi}{dt} \quad (2)$$

These equations lead to equivalent representation of magnetic permeances (inverse reluctances) with capacitances. Hence, the nonlinear permeances of the magnetic circuit paths in Figure 1, are represented as nonlinear capacitances. The permeances can be approximately calculated using the magnetic material properties and the geometric parameters of the magnetic circuit with (3):

$$\rho = \frac{\mu_r \mu_0 l}{A} \quad (3)$$

$\mu_0$= 4π×$10^{-7}$– magnetic permeability of free air
$\mu_r$ – relative magnetic permeability of the core
$A$ – cross-sectional area
$l$ – mean length of the path

## IV. Simulations and Results

The G-C model was built in MATLAB/SIMULINK®. All the simulations were conducted using the one-line diagram shown in Figure 3, with the device and load parameters as listed in Table I. The assumed ferromagnetic core material is M36.

The G-C circuit model is based on the configuration of the CVSR in Figure 1. The control side is connected to an ideal dc source, while the ac circuit is connected in series with a voltage source and a load. Here, gyrators represent the windings. The fringing flux around the air gap has been taken into the account within its permeance [15, 10]. First, the CVSR is simulated under normal condition and steady state for different values of the ac voltage sources and dc bias currents. The behavior of the CVSR in transient conditions and its impact on the system have been addressed in [16].

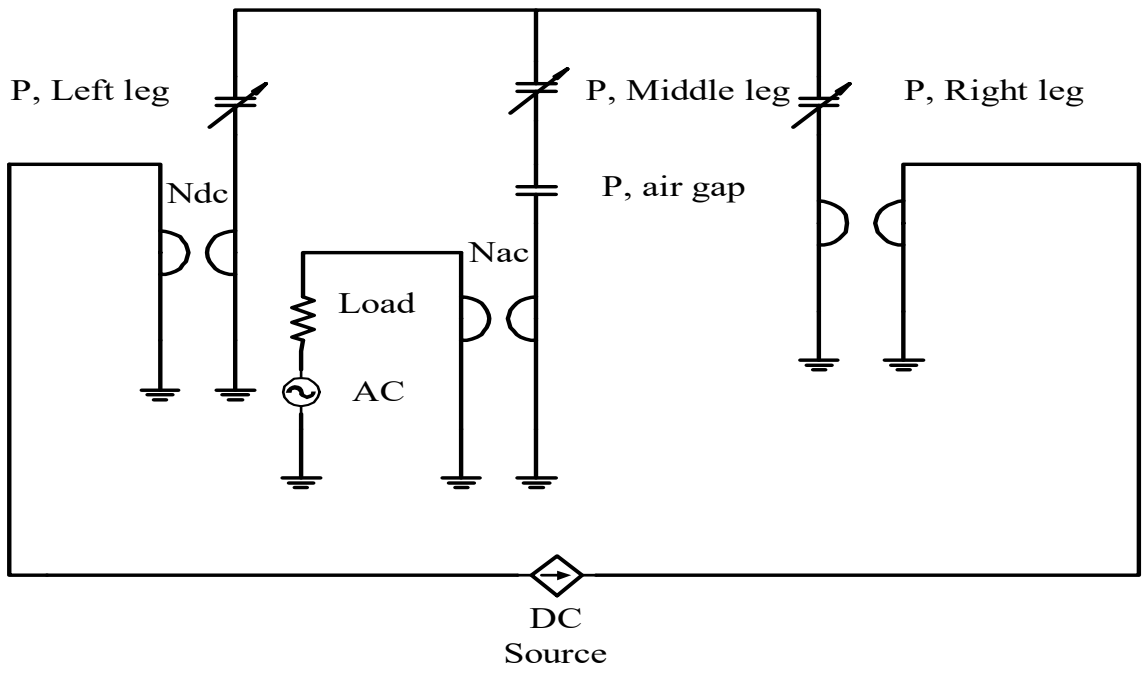

Fig. 3. Gyrator-Capacitor model in MATLAB /SIMULINK®

### A. AC voltage source at 1.2 kV

In this scenario, three different dc bias currents are applied in the control circuit: 0 A, 200 mA (the critical current, as explained below), and 10A.

Figures 4 and 5 represent the flux densities and terminal voltages for 0 A dc bias, respectively. Since the flux densities are purely sinusoidal, the terminal voltage will also be purely sinusoidal. The flux densities through the outer legs are equal at all times. The induced voltages in the two dc windings cancel each other and the equivalent voltage across both is zero.

Table I – CVSR Parameters

| Parameter | Description | Value |
|---|---|---|
| $l_m$ | mean length of the middle leg | 45.72 cm |
| $l_{out}$ | mean length of the outer legs | 86.36 cm |
| $h_{ag}$ | height of the air gap | 0.2014 cm |
| $A$ | cross-section area of the core | 0.0103 $m^2$ |
| $N_{dc}$ | number of turns in the dc winding | 225 |
| $N_{ac}$ | number of turns in the ac winding | 150 |
| $V$ | voltage source | 1.2-3.8 kV |
| $R$ | load resistance | 100 Ω |
| $L$ | load inductance | 130 mH |
| $PF$ | power factor | 0.9 |
| $B_{sat}$ | saturation point | 1.34 T |

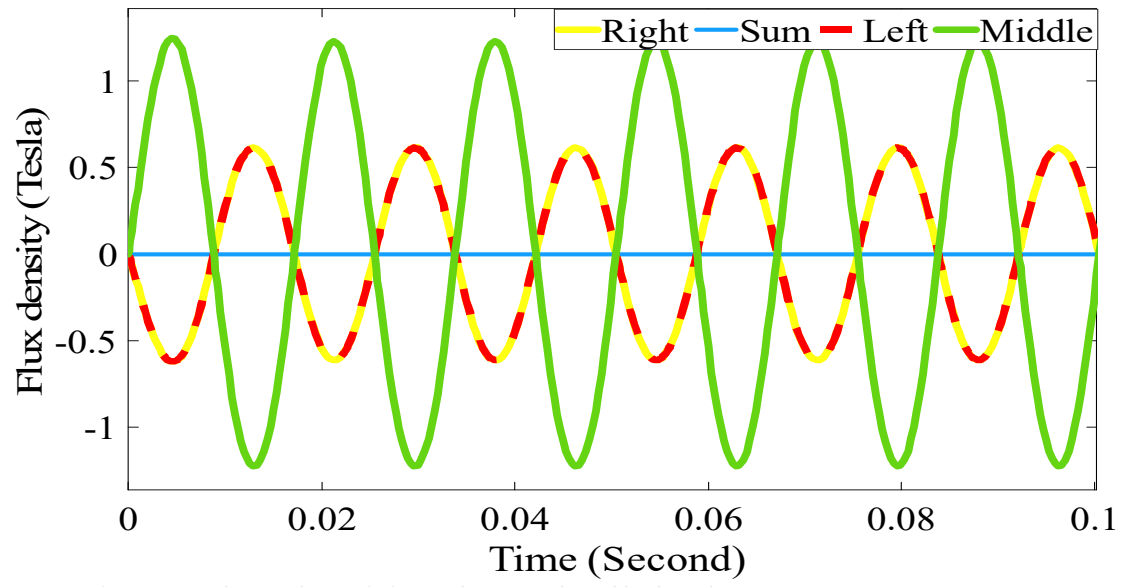

Fig. 4. Flux densities through all the legs ($I_{dc}$ = 0 A)

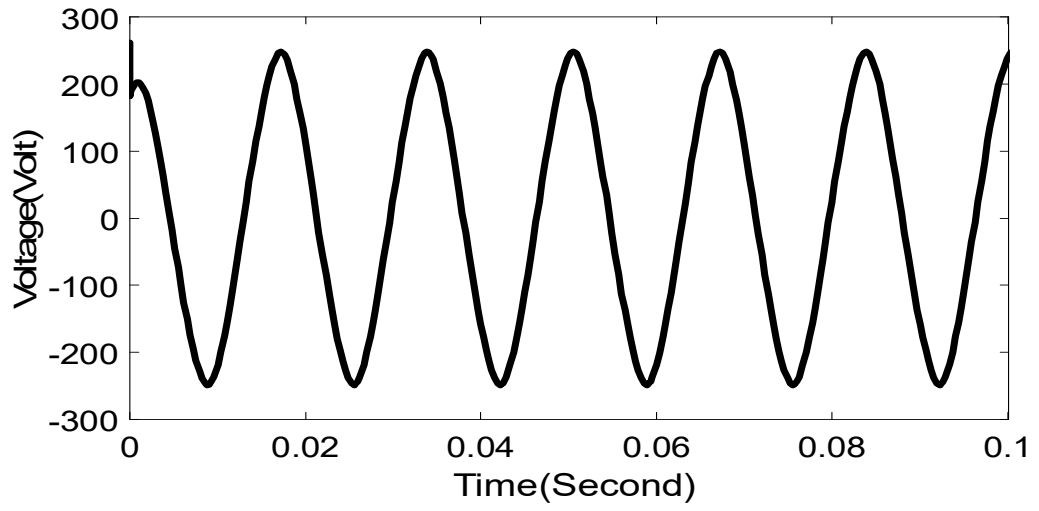

Fig. 5. Induced voltage across the ac winding ($I_{dc}$ = 0 A)

Depending on the region of operation (no saturation, partial saturation, or complete saturation), the equivalent inductance of the CVSR will change. The inductance can be expressed by (4):

$$L = \frac{N_{ac}^2}{R_m} \quad (4)$$

$R_m$ – equivalent reluctance of the device

The equivalent circuit on the electric side, with the voltage source, the equivalent load, and the variable equivalent inductance of the device in series, is as shown in Figure 6. The voltage across the L-CVSR is the terminal voltage equal to the gyrator induced voltage on the primary side based on the G-C model definition.

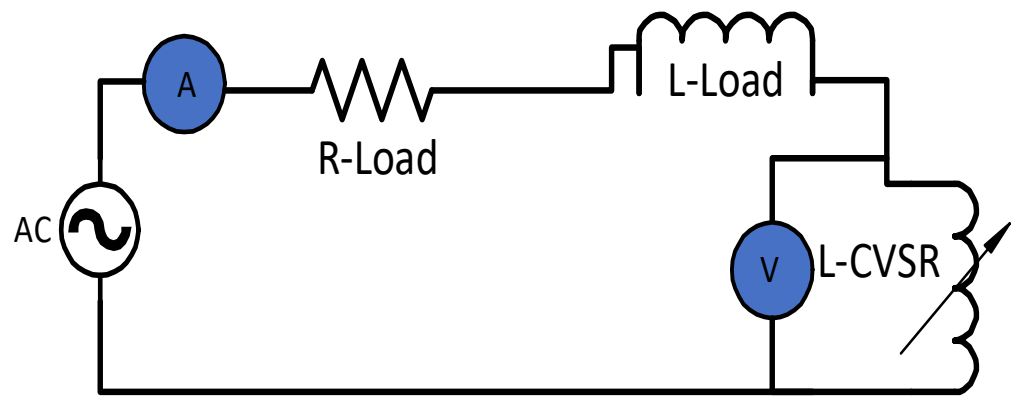

Fig. 6. Electric circuit with induced voltage across the ac winding

The instantaneous and mean value of the equivalent inductance of the device are shown in Figure 7, and the current through the ac winding is shown in Figure 8. Because the device operates in the linear region in this case, inductance changes are negligible (note that the y-axis scale in Figure 7 is zoomed in).

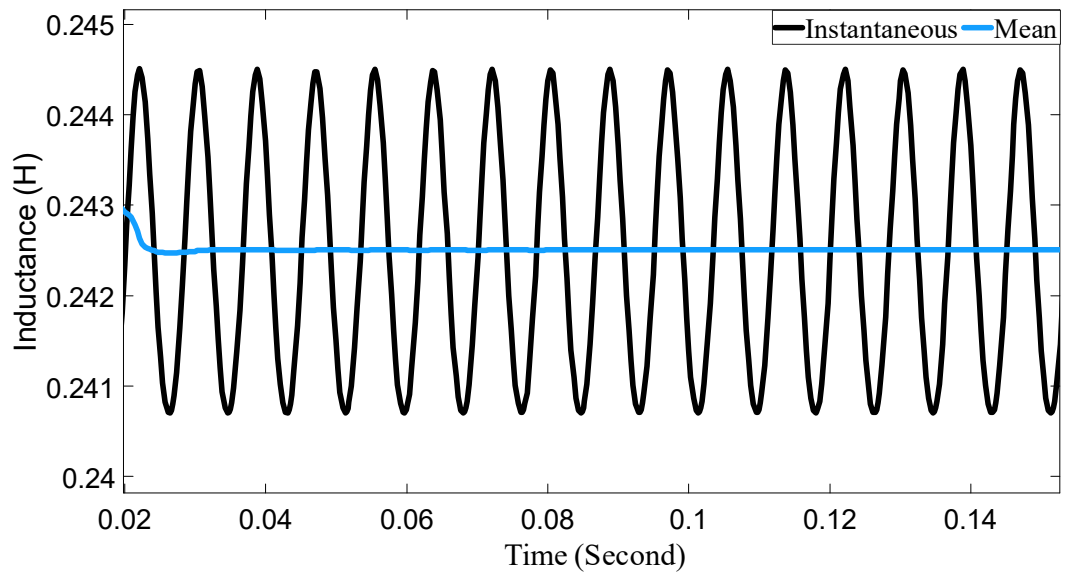

Fig. 7. Instantaneous and mean inductance of the CVSR ($I_{dc}$ = 0 A

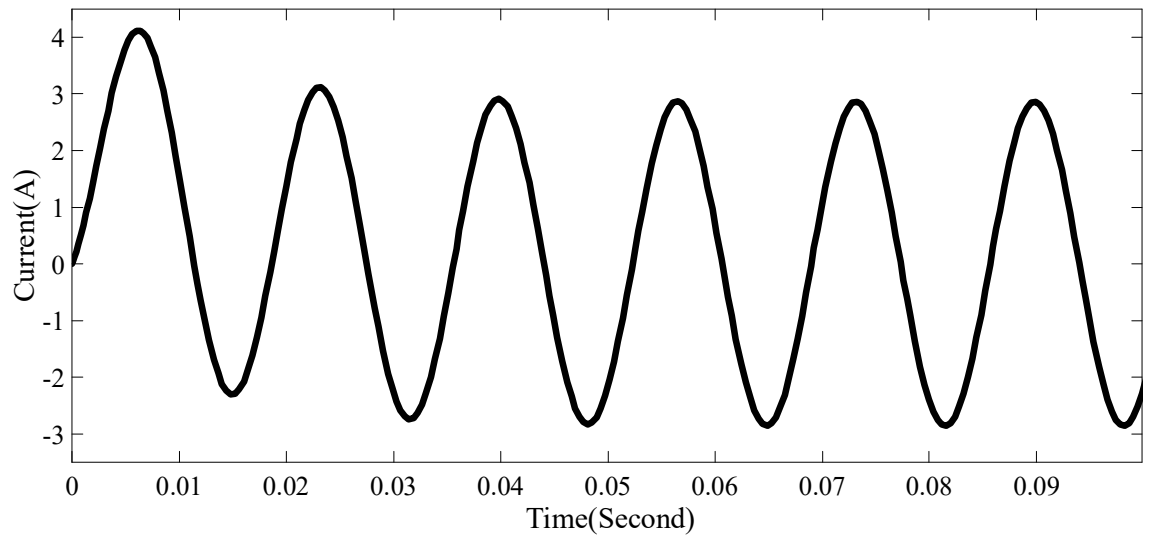

Fig. 8. Current in the ac winding ($I_{dc}$ = 0 A)

With zero dc bias, the power transferred into the dc windings is also zero. The imposed induced voltage in the dc circuit, $E_{emf}$, is proportional to the difference between the flux rates of the outer legs as in (5):

$$E_{emf} = N_{dc}.\left(\frac{d\Phi_{right}}{dt} - \frac{d\Phi_{left}}{dt}\right) \quad (5)$$

As noted above, because the ac flux is divided equally between the outer legs, the induced voltage across the dc winding is zero. Figure 9 shows the flux densities for a dc bias of 200 mA. At this critical current, at all times, one of the outer legs goes into saturation and the other one is unsaturated. Hence, the induced voltage across the ac winding and ac current will change and have some distortion as can be seen on Figures 10 and 11, respectively. The induced voltage across the dc windings will not be zero and has twice the frequency of the system, as shown in Figure 12.

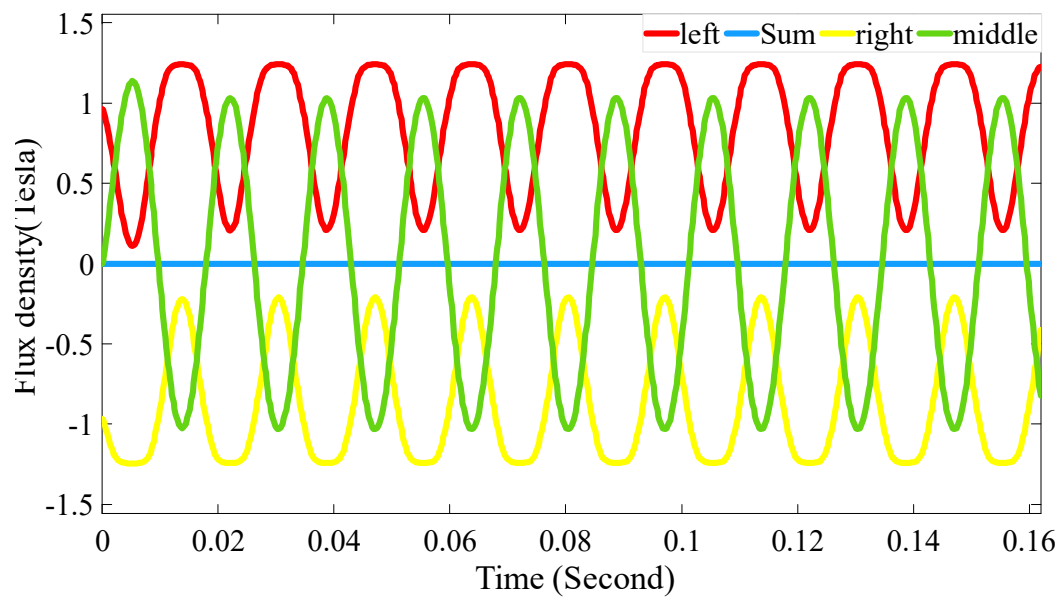

Fig. 9. Flux densities through all the legs ($I_{dc}$ = 200 mA)

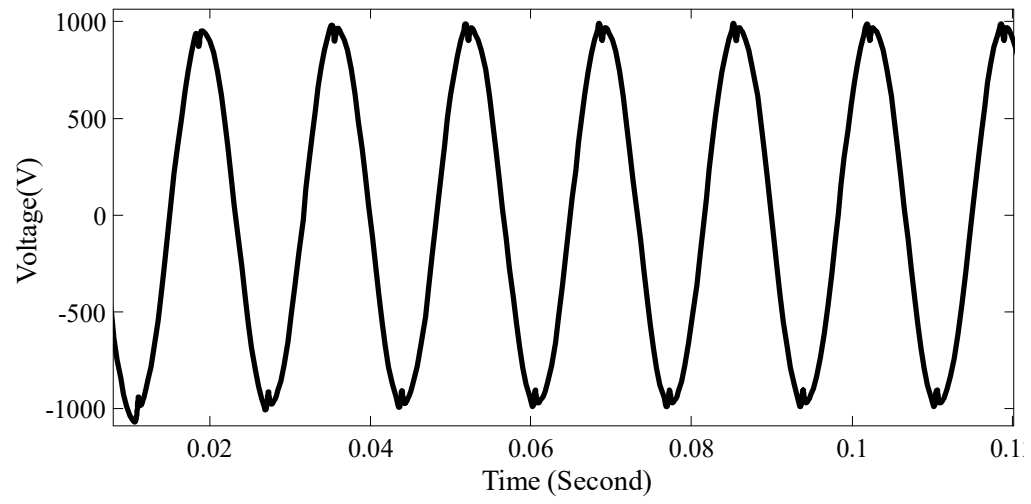

Fig. 10. Induced voltage across the ac winding ($I_{dc}$ = 200 mA)

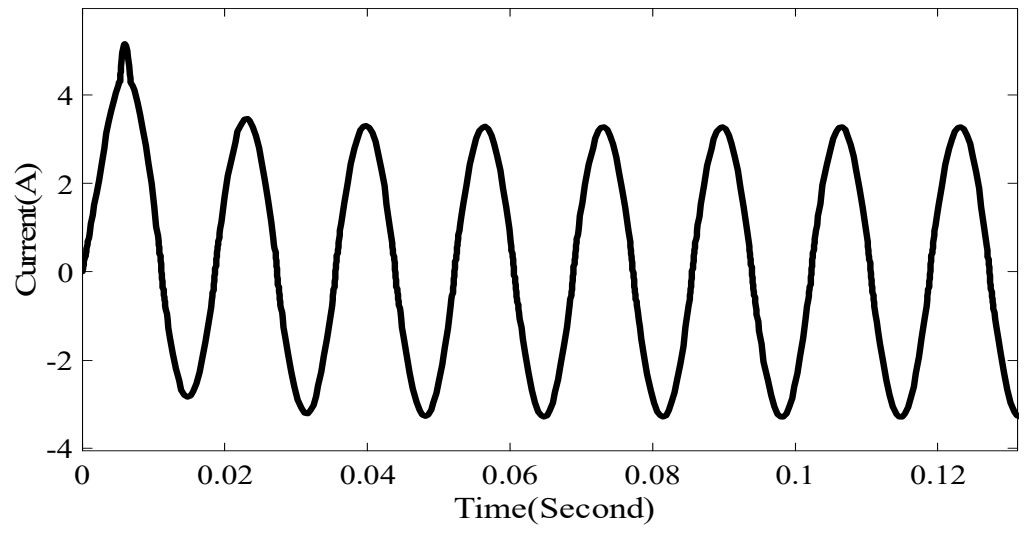

Fig. 11. Current in the ac winding ($I_{dc}$ = 200 mA)

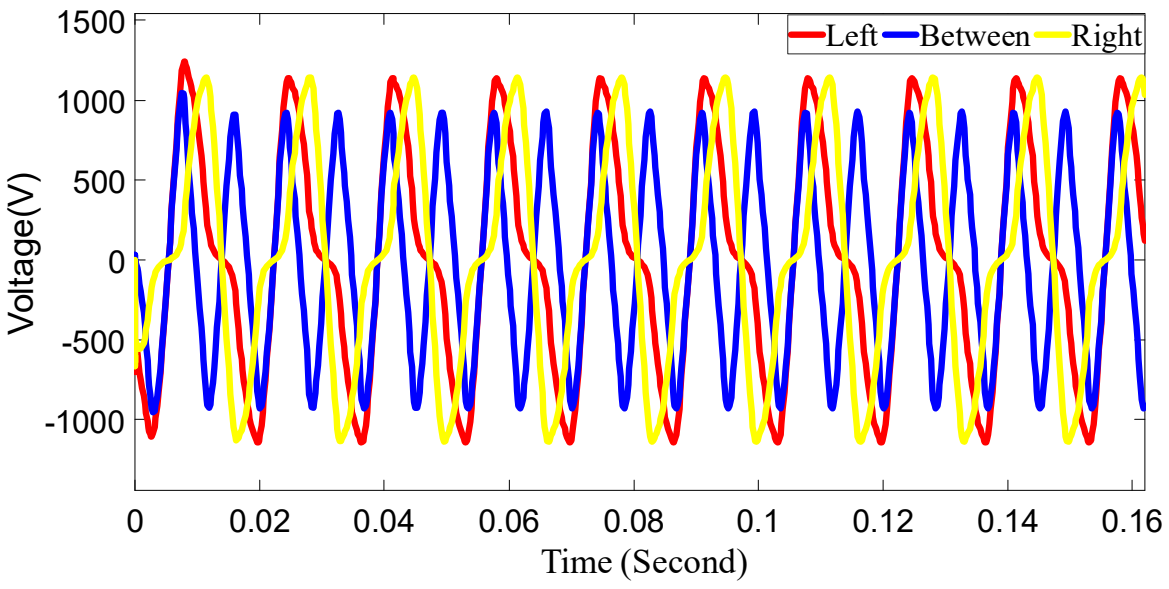

Fig. 12. Induced voltage across the dc windings ($I_{dc}$ = 200 mA)

The instantaneous and the mean value of the equivalent inductance of the device is shown in Figure 13. It can be seen that the peak inductance reaches 0.24 H, equal to the inductance when it operates in the unsaturated region. The mean value is much lower, at about 0.05 H, because the inductance has a small value for a much longer time than it has the peak value.

Figure 14 shows the power transferred into the dc controlled circuit. It can be seen that there is about 2.3 kvar of reactive power and no real power transferred, as expected from a purely reactive element.

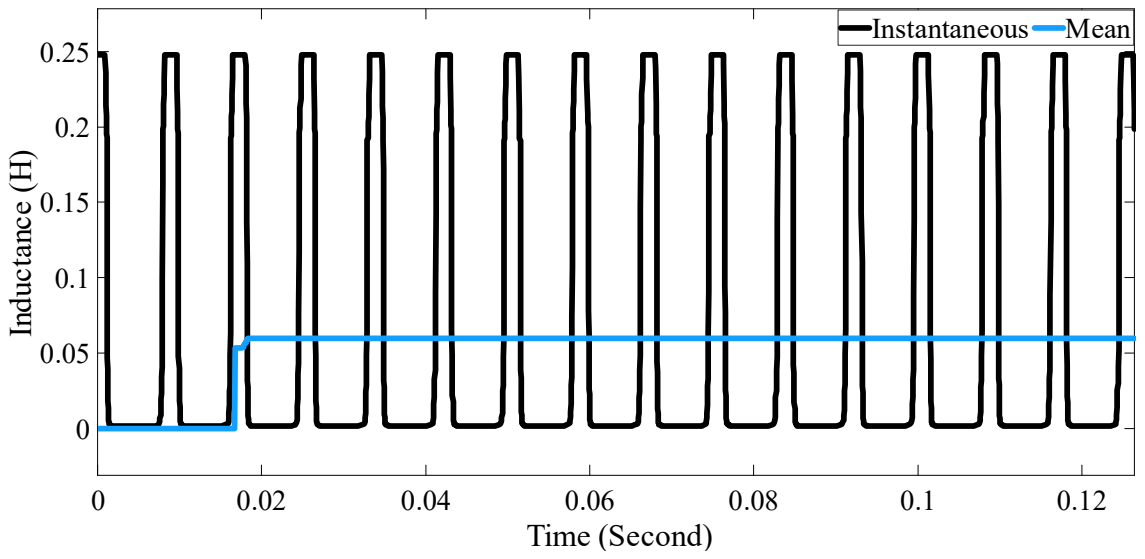

Fig. 13. Instantaneous and mean inductance of the CVSR ($I_{dc}$ = 200 mA)

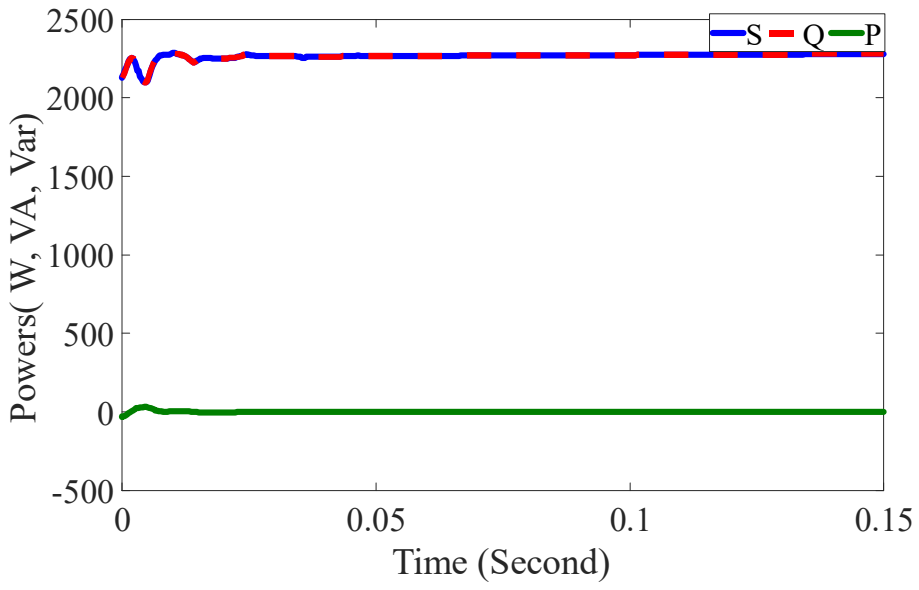

Fig. 14. Power transferred into the dc winding ($I_{dc}$ = 200 mA)

In Figure 15, the CVSR completely goes into saturation because of a high dc offset. The flux density through the middle leg decreased due to the high reluctance of the outer legs.

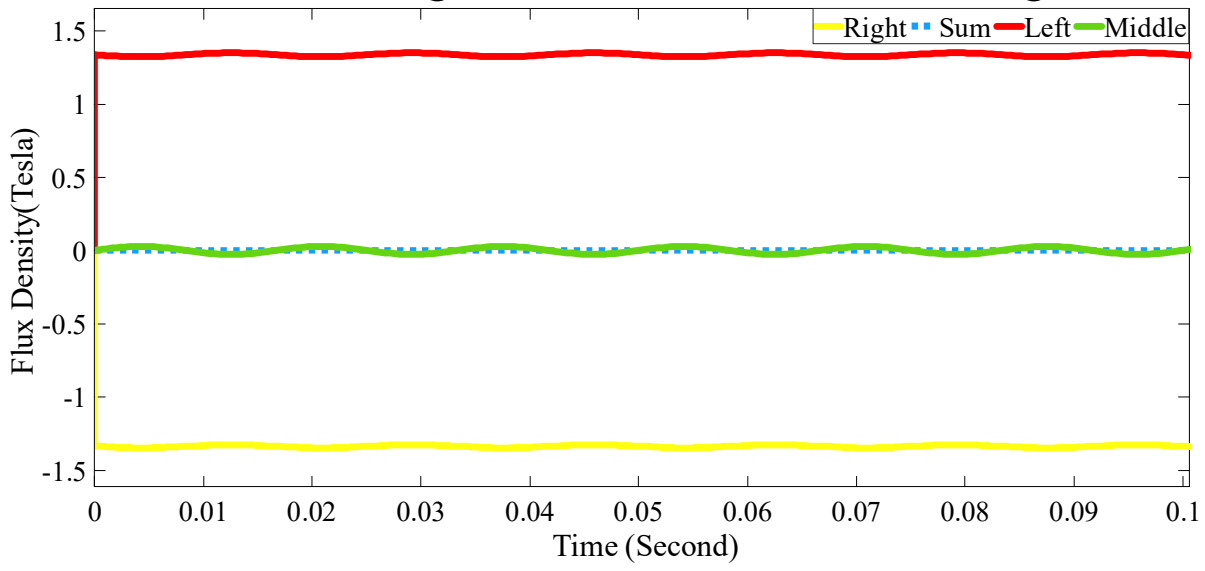

Fig. 15. Flux densities through all the legs ($I_{dc}$ = 10 A)

Current through the ac winding is shown in Figure 16. Due to the complete saturation of the core, reactance in the ac circuit is negligible, so this current is almost equal to the load current.

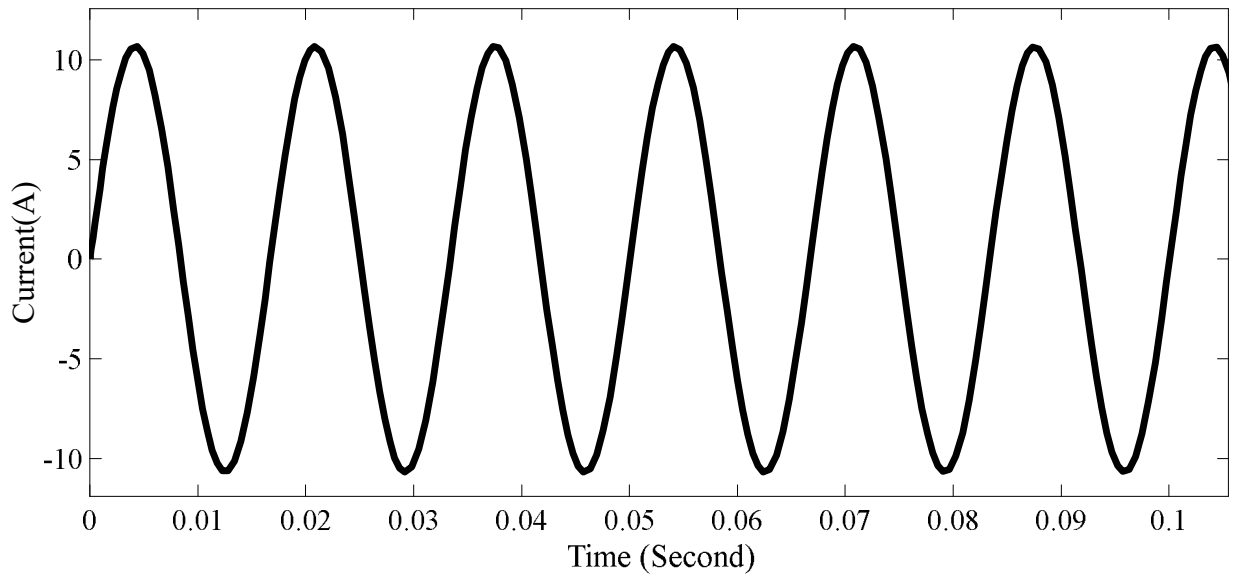

Fig. 16. Current in ac winding ($I_{dc}$ = 10 A)

## B. *AC voltage source at 3.8 kV*

In this scenario, the voltage source is increased to 3.8 kV, causing the core, including the middle leg, to go back and forth into the saturation region, which creates asymmetry and harmonics on both the dc and the ac side. With dc bias at 0 A, Figures 17, 18, and 19, show the flux densities, terminal ac voltage, and the current in the ac winding, respectively. The induced voltage across both dc windings will be zero since the flux densities in the outer legs are again equal at all times. The flat parts in the flux densities are due to the saturation of the core. As a result, the terminal voltage and ac current are distorted at those time instances.

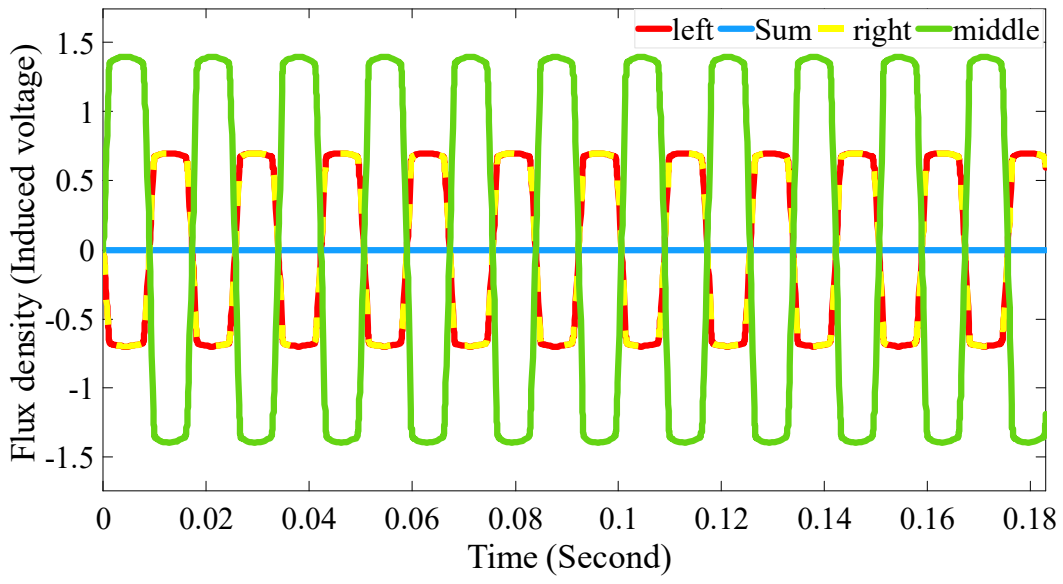

Fig. 17. Flux densities through all the legs ($I_{dc}$ = 0 A)

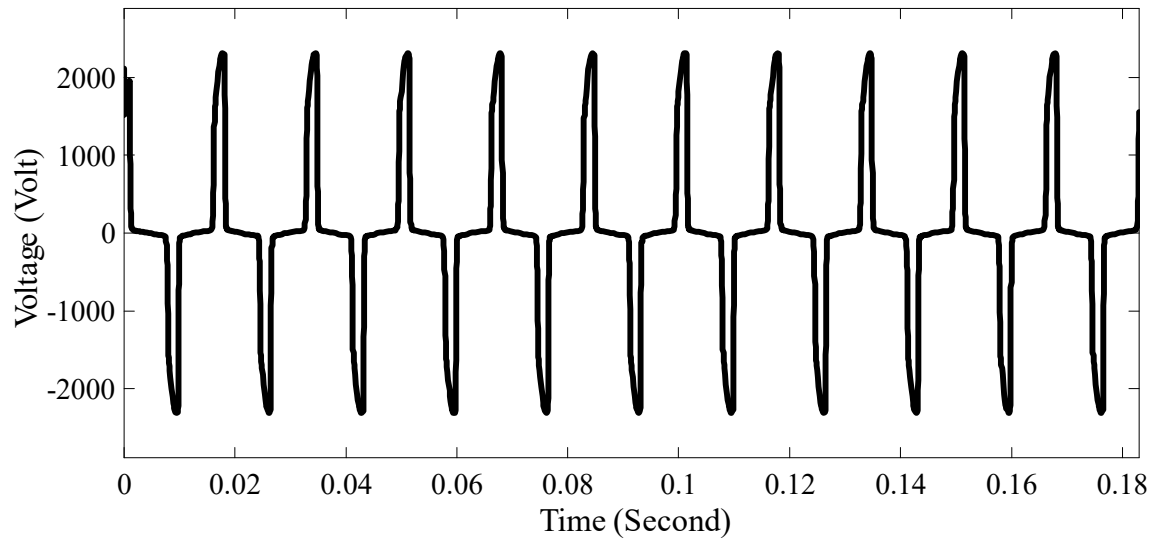

Fig. 18. Induced voltage across the ac winding ($I_{dc}$ = 0 A)

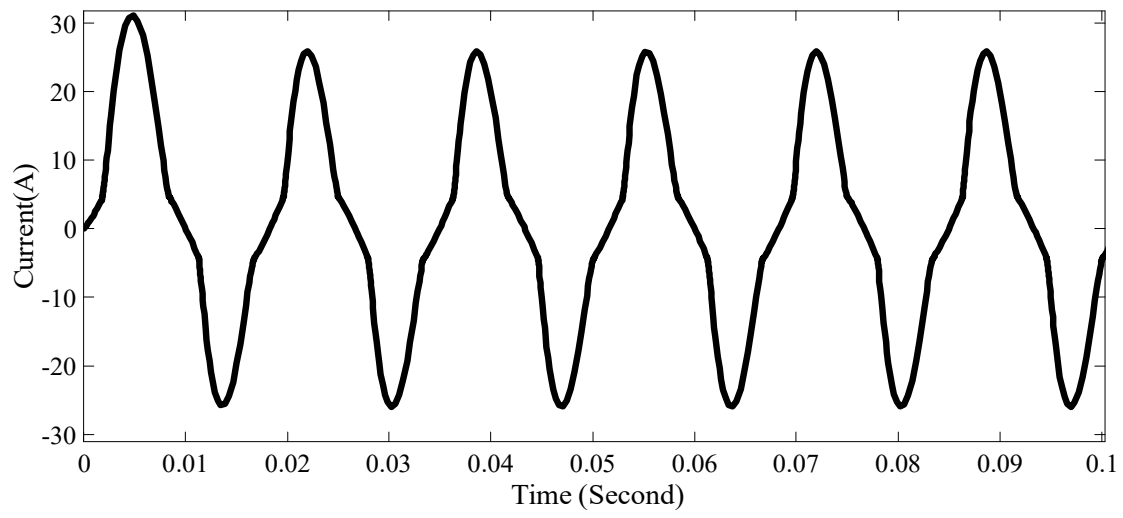

Fig. 19. Current in ac winding ($I_{dc}$ = 0 A)

In Figures 20 and 21, the flux densities and the induced voltage across both dc windings are shown for dc bias at 200 mA. It can be seen that the induced voltage across dc windings has again frequency twice the frequency of the ac circuit. The induced voltage across the ac winding and the ac current have the same peaks as in the previous case (0 A dc). As in the first scenario ($V_{ac}$ = 1.2 kV), they will be distorted in this case.

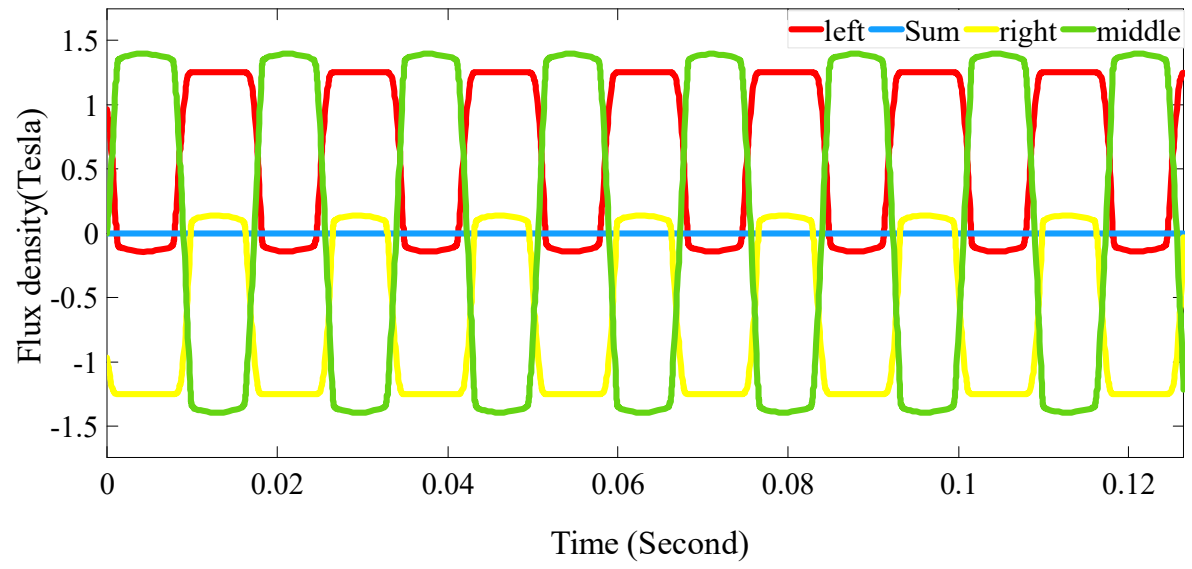

Fig. 20. Flux densities through all the legs ($I_{dc}$ = 200 mA)

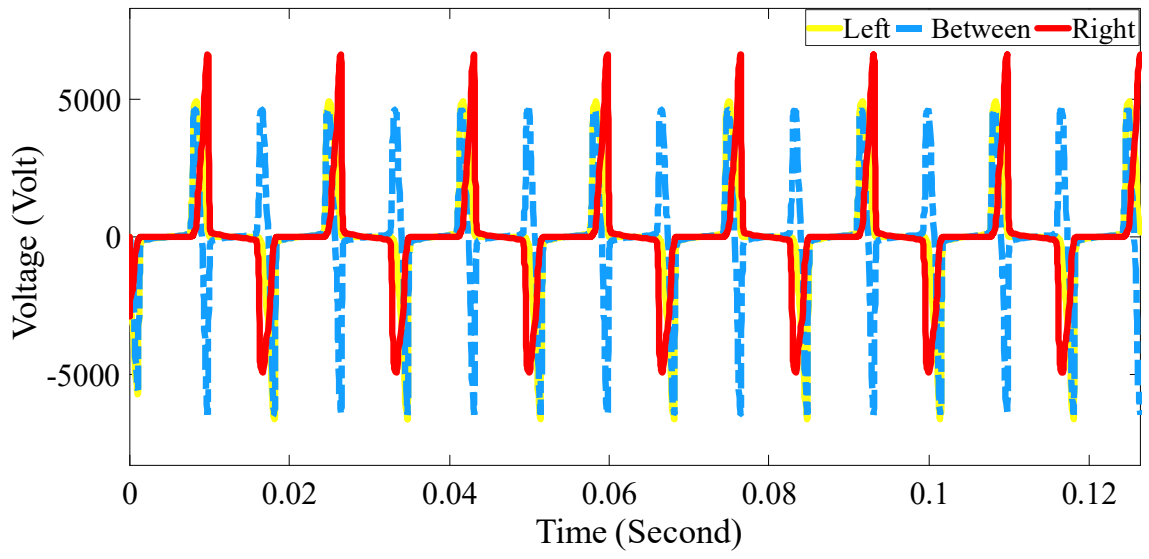

Fig. 21. Induced voltage across the dc windings ($I_{dc}$ = 200 mA)

Furthermore, since the device does not completely go into saturation, the mean equivalent inductance of the device for both cases (0 and 200 mA) is almost the same. It can be seen that the induced voltage across dc windings is not symmetrical.

In Figure 22, the current through the ac winding is shown for dc bias at 10A. Since the bias is very high, the core is deeply saturated. The flux density through all the legs are similar to the first scenario ($V_{ac}$ = 1.2 kV), at 10A dc bias (Figure 15).

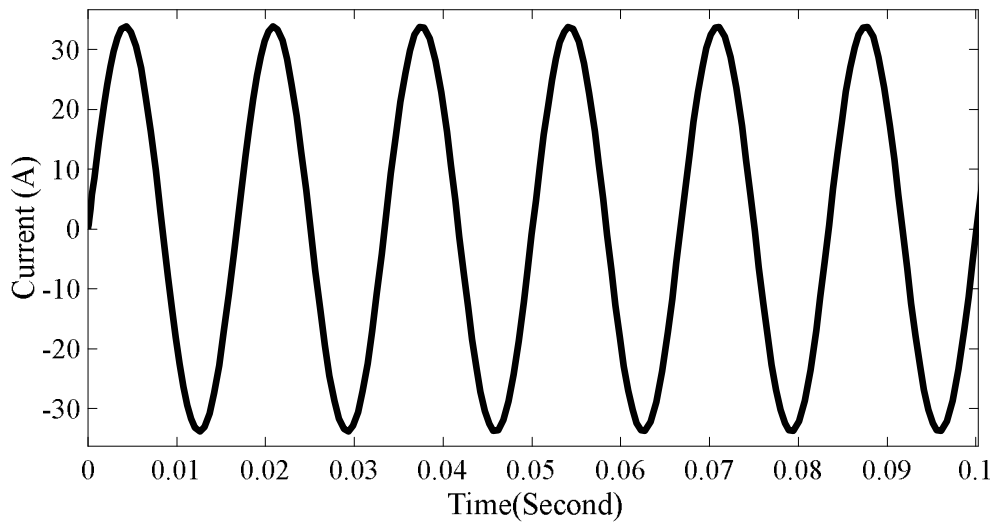

Fig. 22. Current in ac winding ($I_{dc}$ = 10 A)

## V. CONCLUSION

The paper presents results from a simulation study of CVSR operation in different steady state modes. The focus is on induced voltages across the dc windings, flux densities through the core legs, and the power transferred into the dc circuit.

A G-C model which keeps power equivalence is used to provide the analogy between the magnetic and the electric circuit. The model is based on the configuration of a three-legged CVSR. Capacitors, equivalent to permeances, are used to model core characteristics. Simulations were carried out using different values of the bias dc current and the ac voltage source for a comprehensive analysis of the CVSR behaviour. Future work will investigate a more complicated and realistic model of the device and its behavior during transient conditions.

## REFERENCES

[1] A. F. Soofi, S. D. Manshadi, G. Liu and R. Dai, "A SOCP Relaxation for Cycle Constraints in the Optimal Power Flow Problem," in IEEE Transactions on Smart Grid, vol. 12, no. 2, pp. 1663-1673, March 2021.

[2] Taghavirashidizadeh A, Parsibenehkohal R, Hayerikhiyavi M, Zahedi M. A Genetic algorithm for multi-objective reconfiguration of balanced and unbalanced distribution systems in fuzzy framework. Journal of Critical Reviews. 2020;7(7):639-343.

[3] A. Dimitrovski, Z. Li, and B. Ozpineci, "Applications of saturable-core reactors (SCR) in power systems," presented at the IEEE Power Energy Soc. T&D Conf. Expo., Chicago, IL, USA, 2014.

[4] A. Dimitrovski, Z. Li, and B. Ozpineci, "Magnetic amplifier-based power flow controller," IEEE Trans. Power Del., vol. 30, no. 4, pp. 1708–1714, Aug. 2015.

[5] S. Khazaee, M. Hayerikhiyavi, S. M. Kouhsari, "A Direct-Based Method for Real-Time Transient Stability Assessment of Power Systems", CRPASE Vol. 06(02), 108-113, June 2020.

[6] N. S. Gilani, M. T. Bina, F. Rahmani and M. H. Imani, "Data-mining for Fault-Zone Detection of Distance Relay in FACTS-Based Transmission," 2020 IEEE Texas Power and Energy Conference (TPEC), College Station, TX, USA, 2020, pp. 1-6.

[7] H. Haggi, S. R. Marjani and M. A. Golkar, "The effect of rescheduling power plants and optimal allocation of STATCOM in order to Improve power system static security using TLBO algorithm," *2017 Iranian Conference on Electrical Engineering (ICEE)*, Tehran, 2017, pp. 1120-1125.

[8] F. Moriconi, F. De La Rosa, F. Darmann, A. Nelson, and L. Masur, "Development and deployment of saturated-core fault current limiters in distribution and transmission substations," *IEEE Trans. Appl. Superconductivity*, vol. 21, no. 3, pp. 1288–1293, Jun. 2011.

[9] T. Wass, S. Hornfeldt, and S. Valdemarsson, "Magnetic circuit for a controllable reactor," IEEE Transactions on Magnetics, vol. 42, pp. 2196-2200, 2006.

[10] M. Young, A. Dimitrovski, Z. Li and Y. Liu, "Gyrator-Capacitor Approach to Modeling a Continuously Variable Series Reactor," *IEEE Trans. on Power Delivery*, vol. 31, no. 3, pp. 1223-1232, June 2016.

[11] M. Young, Z. Li and A. Dimitrovski, "Modeling and simulation of continuously variable series reactor for power system transient analysis," *2016 IEEE Power and Energy Society General Meeting (PESGM), Boston*, MA, 2016, pp. 1-5.

[12] L. Yan and B. Lehman, "Better understanding and synthesis of integrated magnetics with simplified gyrator model method," *2001 IEEE 32nd Annual Power Electronics Specialists Conference (IEEE Cat. No.01CH37230),* Vancouver, BC, 2001, pp. 433-438 vol. 1.

[13] D. C. Hamill, "Gyrator-capacitor modeling: a better way of understanding magnetic components," *Proceedings of 1994 IEEE Applied Power Electronics Conference and Exposition - ASPEC'94*, Orlando, FL, USA, 1994, vol.1, pp. 326-332.

[14] E. Rozanov, S. Ben-Yaakov," Analysis of Current-controlled Inductors by New SPICE Behavioral Model," HAIT Journal of Science and Engineering, vol. 2, pp. 558-570.

[15] S. D. Sudhoff, "Magnetics and Magnetic Equivalent Circuits," in Power Magnetic Devices: A Multi-Objective Design Approach. Wiley-IEEE Press, 2014, pp. 45–112.

[16] Hayerikhiyavi M, Dimitrovski A. "Comprehensive Analysis of Continuously Variable Series Reactor Using GC Framework". arXiv preprint arXiv:2103.11136. 2021 Mar 20.